\documentclass[12pt,a4paper]{article} 

\usepackage{epsfig} 
\usepackage{amssymb} 
\usepackage{amsmath} 
\usepackage{amsbsy} 
\usepackage{bm} 
\usepackage{graphicx}
\usepackage{url}
\usepackage[round]{natbib}
\usepackage{setspace}
\usepackage[margin=2.5cm]{geometry}
\usepackage[max2]{authblk}
\def\Levy{L\'{e}vy}

\begin{document} 
\bibliographystyle{plainnat}
\renewcommand{\thetable}{\Roman{table}}
\renewcommand{\thefigure}{\arabic{figure}}

\renewcommand\Affilfont{\itshape\small}

\title{Parameter estimation of a \Levy\ copula of a discretely observed bivariate compound Poisson process with an application to operational risk modelling} 
\author[1,2]{J.\ L. van Velsen%
\thanks{Views expressed in the paper are the author's own and do not necessarily reflect those of ABN Amro. Electronic address: \texttt{joris.van.velsen@nl.abnamro.com}}} 
\affil[1]{EC Modelling, ABN Amro, P.O. Box 283, 1000 EA Amsterdam, The Netherlands}
\affil[2]{Duisenberg school of finance, Gustav Mahlerplein 117, 1082 MS Amsterdam, The Netherlands}
\date{August 2012}
\maketitle 

\begin{abstract}
A method is developed to estimate the parameters of a \Levy\ copula of a discretely observed bivariate compound Poisson process without knowledge of common shocks. The method is tested in a small sample simulation study. Also, the method is applied to a real data set and a goodness of fit test is developed. With the methodology of this work, the \Levy\ copula becomes a realistic tool of the advanced measurement approach of operational risk.  
\end{abstract}

\section{Introduction}

The multivariate compound Poisson process is an intuitively appealing and natural model for operational risk and insurance claim modelling. The model is
intuitively appealing because dependencies between different loss categories are caused by common shocks that apply to multiple loss categories simultaneously. For example, in operational risk modelling, failure of an IT system is a common shock that causes losses in multiple lines of business. The multivariate compound Poisson process is a natural model for the following two reasons. First, as a L\'{e}vy process, it is easily applied to any time horizon of interest. Second, because a redesign of loss categories results in a loss process that is again multivariate compound Poisson, the nature of the model does not depend on the level of granularity \citep{boc08}. 

A multivariate compound Poisson process can be specified in terms of univariate compound Poisson processes and a \Levy\ copula \citep{cont04}. 
In essence, a \Levy\ copula provides the relationship between the \Levy\ measure of a multivariate \Levy\ process and the \Levy\ measures of its marginal processes. The \Levy\ copula allows for a parsimonious bottom-up modelling with compound Poisson processes. In case of two loss categories, for example, parameterization with a Clayton \Levy\ copula requires two marginal frequencies, two marginal jump size distributions and one \Levy\ copula parameter. In contrast, parameterization without a \Levy\ copula requires three frequencies (corresponding to losses of the first category only, losses of the second category only and common shocks that apply to both categories), two univariate jump size distributions (corresponding to losses of one of the two categories only) and, finally, one bivariate jump size distribution (corresponding to the common shocks).     

The parameters of a \Levy\ copula of a multivariate compound Poisson process can be estimated if the process is either observed continuously (such that common shocks can be identified) or observed discretely with knowledge about all jump sizes and the common shocks. These two cases have been studied by \citet{Esm01} for a bivariate compound Poisson process (the continuous observation is mimicked in a simulation study, while the discrete observation corresponds to a real data set of insurance claims). The objective of this work is to develop a method to estimate the parameters of a \Levy\ copula of a bivariate compound Poisson process in case the process is observed discretely with knowledge about all jump sizes, but without knowledge of which jumps stem from common shocks. This situation is relevant to operational risk modelling in which all material losses are registered, but common shocks are typically unknown.  
With the methodology developed here, the \Levy\ copula becomes a realistic tool of the advanced measurement approach of operational risk. 

The outline of this paper is as follows. In Section \ref{CPP}, we discuss the bivariate compound Poisson process in terms of its common shock representation and \Levy\ measure. This prepares the ground for the two-dimensional \Levy\ copula of Section \ref{levy_copula}. In Section \ref{MLE}, the new estimation method of the discretely observed bivariate compound Poisson process is presented. The method is tested in a simulation study in Section \ref{sim_study} and applied to a real data set in Section \ref{real_data}. In Section \ref{real_data}, we also develop a goodness of fit test for the \Levy\ copula. Finally, we conclude in Section \ref{conclusions}.

\section{The bivariate compound Poisson process}
\label{CPP}

A bivariate compound Poisson process $S=(S(t))_{t \ge 0}=((S_{1}(t),S_{2}(t)))_{t \ge 0}$ is defined on a filtered probability space $(\Omega,\mathcal{F},\mathbb{P})$ as
\begin{equation}
S(t)=\sum_{i=1}^{N(t)} Y_{i},
\label{cpp_def}
\end{equation}
where $N=(N(t))_{t \ge 0}$ is a Poisson process with intensity $\lambda>0$ and $Y=(Y_{i})_{i \in \mathbb{N}^{+}}$ is a sequence of iid $2$-dimensional random vectors. The process $N$ and the sequence $Y$ are statistically independent. By construction, given any $t>s$, the increment $S(t)-S(s)$ is independent of $\mathcal{F}(s)$ and has the same distribution as $S(t-s)$. The probability distribution of $Y_{i}$ is such that $\mathbb{P}(Y_{i}=0)=0$, which means that a jump of $N$ almost surely manifests itself in a jump of at least one of the components of $S$.    

\subsection{The L\'{e}vy-It\^{o} decomposition}

The L\'{e}vy-It\^{o} decomposition of $S(t)$ takes the form 
\begin{equation}
S(t)=\int_{0}^{t} \int_{\mathbb{R}^{2} \backslash \{0\}} xJ(ds,dx),
\label{levy_ito}
\end{equation}  
where $J$ is the Poisson random measure. With the help of the L\'{e}vy-It\^{o} decomposition, we find that $S(t)$ has common shock representation
\begin{equation}
S_{1}(t)=S_{1}^{\perp}(t)+S_{1}^{\parallel}(t) \quad \mbox{and} \quad S_{2}(t)=S_{2}^{\perp}(t)+S_{2}^{\parallel}(t), 
\end{equation}    
where 
\begin{equation}
S_{1}^{\perp}(t)=\int_{0}^{t} \int_{\mathbb{R} \backslash \{0\}} x_{1}J(ds,dx_{1} \times \{0\}),
\label{S1perp}
\end{equation}
\begin{equation}
S_{2}^{\perp}(t)=\int_{0}^{t} \int_{\mathbb{R} \backslash \{0\}} x_{2}J(ds,\{0\} \times dx_{2})
\label{S2perp}
\end{equation}
and
\begin{equation}
S^{\parallel}(t) \equiv (S_{1}^{\parallel}(t),S_{2}^{\parallel}(t))=\int_{0}^{t} \int_{(\mathbb{R} \backslash \{0\})^{2}} (x_{1},x_{2}) J(ds,dx_{1} \times dx_{2}).
\label{Sparallel}
\end{equation}
The processes $S_{1}^{\perp}=(S_{1}^{\perp}(t))_{t \ge 0}$, $S_{2}^{\perp}=(S_{2}^{\perp}(t))_{t \ge 0}$ and $S^{\parallel}=(S^{\parallel}(t))_{t \ge 0}$ do not jump simultaneously and are statistically independent \citep{sat99}. The processes $S_{1}^{\perp}$ and $S_{2}^{\perp}$ are called the independent parts of $S$. Conversely, the process $S^{\parallel}$ is called the dependent part of $S$ and corresponds to the common shocks. 

\subsection{The L\'{e}vy-Khinchin representation and the \Levy\ measure}

The L\'{e}vy-Khinchin representation of the characteristic function of $S(t)$ can be determined from Eq.\ (\ref{levy_ito}) with the exponential formula for Poisson random measures \citep{cont04}. The representation takes the form
\begin{equation}
\mathbb{E} \left[ e^{i u \cdot S(t)} \right]=\exp{ \left[t \int_{\mathbb{R}^{2}}\left(e^{i u \cdot x}-1\right)\nu(dx) \right]}, 
\label{char}
\end{equation}
where $u \in \mathbb{R}^{2}$ and the L\'{e}vy measure
\begin{equation}
\nu(A)=\lambda \mathbb{P}(Y_{i} \in A)=\frac{\mathbb{E}[J((0,t],A)]}{t}
\end{equation}
gives the expected number of jumps per unit of time in each Borel set $A$ of $\mathbb{R}^{2}$. The processes $S_{1}$ and $S_{2}$ are independent if and only if the support of $\nu$ is contained in the set $\{ x \in \mathbb{R}^{2}: x_{1}x_{2}=0 \}$ \citep{cont04}. In this case, $S_{1}$ and $S_{2}$ do not jump simultaneously almost surely and the L\'{e}vy-Khinchin representation factorizes as
\begin{equation}
\mathbb{E} \left[ e^{iu_{1}S_{1}(t)+iu_{2}S_{2}(t)} \right] = \mathbb{E} \left[ e^{iu_{1}S_{1}(t)} \right]
\mathbb{E} \left[ e^{iu_{2}S_{2}(t)} \right]. 
\end{equation}
On the other hand, the processes $S_{1}$ and $S_{2}$ are defined to be comonotonic if their jump sizes $\Delta S_{1}$ and $\Delta S_{2}$, respectively, are elements of an increasing set $\mathcal{S}$ of $\mathbb{R}^{2}$, see \citep{cont04}. Any two elements $(x_{1},x_{2})$ and $(y_{1},y_{2})$ of $\mathcal{S}$ satisfy $x_{i}>y_{i}$ or $x_{i}<y_{i}$ for all $i=1,2$. An example of an increasing set is $\{ x \in \mathbb{R}^{2}: x_{1}=x_{2}\}$. The requirement $(\Delta S_{1},\Delta S_{2}) \in \mathcal{S}$ means that by observing one of the processes $S_{1}$ or $S_{2}$, the other process can be constructed exactly with a positive dependence. 
In case of comonotonic $S_{1}$ and $S_{2}$, the \Levy\ measure is concentrated on $\mathcal{S}$.  

In terms of $S_{1}^{\perp}(t)$, $S_{2}^{\perp}(t)$ and $S^{\parallel}$, Eq.\ (\ref{char}) takes the form
\begin{equation}
\mathbb{E} \left[ e^{iu_{1}S_{1}(t)+iu_{2}S_{2}(t)} \right] = \mathbb{E} \left[ e^{iu_{1}^{\vphantom{\perp}}S_{1}^{\perp}(t)} \right]
\mathbb{E} \left[ e^{iu_{2}^{\vphantom{\perp}}S_{2}^{\perp}(t)} \right] \mathbb{E} \left[ e^{iu_{1}^{\vphantom{\perp}}S_{1}^{\parallel}(t)+iu_{2}^{\vphantom{\perp}}S_{2}^{\parallel}(t)} \right], 
\label{char_detailed}
\end{equation}
where we have used that $S_{1}^{\perp}$, $S_{2}^{\perp}$ and $S^{\parallel}$ are independent. The L\'{e}vy-Khinchin representation of the characteristic functions of $S_{1}^{\perp}(t)$, $S_{2}^{\perp}(t)$ and $S^{\parallel}(t)$ can be determined from their L\'{e}vy-It\^{o} decompositions in the same way as Eq.\ (\ref{char}) is determined from Eq.\ (\ref{levy_ito}). The \Levy\ measures of $S_{1}^{\perp}$ and $S_{2}^{\perp}$ are given by, respectively, 
\begin{equation}
\nu_{1}^{\perp}(B)=\nu(B \times \{0\}) \quad \mbox{and} \quad \nu_{2}^{\perp}(B)=\nu(\{0\} \times B),  
\end{equation}
where $B$ is a Borel set of $\mathbb{R}$. The Levy measure of $S^{\parallel}$ takes the form
\begin{equation}
\nu^{\parallel}(A)=\nu(A)-\nu(A_{1} \times \{0\})-\nu(\{0\} \times A_{2}),
\end{equation}
where the sets $A_{1}$ and $A_{2}$ are defined as  
\begin{equation}
A_{1}=\{ x \in \mathbb{R}^{2}: (x_{1},0) \in A \} \quad \mbox{and} \quad
A_{2}=\{ x \in \mathbb{R}^{2}: (0,x_{2}) \in A \}.
\end{equation}   
To conclude our discussion of the bivariate compound Poisson process $S$, we consider its components for a \Levy\ measure $\nu$ that is not necessarily concentrated on $\{ x \in \mathbb{R}^{2}: x_{1}x_{2}=0 \}$ or an increasing set $\mathcal{S}$. The process $S_{1}$ is compound Poisson \citep{cont04} and by setting $u=(u_{1},0)$ in Eq.\ (\ref{char_detailed}), we find that the characteristic function of $S_{1}(t)$ takes the form
\begin{equation}
\begin{split}
\mathbb{E} \left[ e^{iu_{1}S_{1}(t)} \right] & = \exp{\left[t\int_{\mathbb{R}}\left( e^{iu_{1}x_{1}}-1 \right)\nu_{1}^{\perp}(dx_{1})\right]}  
\exp{\left[t\int_{\mathbb{R}^{2}}\left( e^{iu_{1}x_{1}}-1 \right)\nu^{\parallel}(dx_{1} \times dx_{2})\right]} \\
& = \exp{\left[t\int_{\mathbb{R}}\left( e^{iu_{1}x_{1}}-1 \right)\left\{\nu_{1}^{\perp}(dx_{1})+\nu^{\parallel}(dx_{1} \times (-\infty,\infty))\right\} \right]}.
\label{char_1}
\end{split}
\end{equation}
From Eq.\ (\ref{char_1}), it follows that the \Levy\ measure of $S_{1}$ is given by
\begin{equation}
\nu_{1}(B)=\nu_{1}^{\perp}(B)+\nu^{\parallel}(B \times \mathbb{R}).
\label{nu_1}
\end{equation}
If the measure $\nu$ is concentrated on $\{ x \in \mathbb{R}^{2}: x_{1}x_{2}=0 \}$, then $\nu_{1}(B)=\nu_{1}^{\perp}(B)$ and if it is concentrated on $(\mathbb{R}\backslash \{0\})^{2}$ (such as on an increasing set $\mathcal{S}$), then $\nu_{1}(B)=\nu^{\parallel}(B \times \mathbb{R})$. In general, $\nu_{1}$ is a combination of $\nu_{1}^{\perp}$ and $\nu^{\parallel}$ cf.\ Eq.\ (\ref{nu_1}). In the same way, $S_{2}$ is compound Poisson with \Levy\ measure
\begin{equation}
\nu_{2}(B)=\nu_{2}^{\perp}(B)+\nu^{\parallel}(\mathbb{R} \times B).
\end{equation}

\section{The \Levy\ copula}
\label{levy_copula}

We consider a bivariate compound Poisson process with positive jumps. This means that the \Levy\ measure is concentrated on $[0,\infty)^{2}\backslash \{0\}$ rather than on $\mathbb{R}^{2}\backslash \{0\}$. The assumption of positive jumps is reasonable in the context of operational risk modelling and restricts our discussion of \Levy\ copulas to positive \Levy\ copulas. 

\subsection{The positive \Levy\ copula and Sklar's theorem}

A two-dimensional positive \Levy\ copula $\mathcal{C}: [0,\infty ]^{2} \rightarrow [0,\infty ]$ is a 2-increasing grounded function with uniform margins. The 2-increasing property means that for any $(u_{1},u_{2}) \in [0,\infty]^{2}$ and $(v_{1},v_{2}) \in [0,\infty]^{2}$ with $u_{1}\ge v_{1}$ and $u_{2} \ge v_{2}$, we have 
\begin{equation}
\mathcal{C}(u_{1},u_{2})-\mathcal{C}(u_{1},v_{2})-\mathcal{C}(v_{1},u_{2})+\mathcal{C}(v_{1},v_{2}) \ge 0.
\end{equation}
The grounded property means that $\mathcal{C}(u_{1},u_{2})=0$ if $u_{1}=0$ and/or $u_{2}=0$. Finally, margins are defined as $\mathcal{C}(u_{1},\infty)$ and $\mathcal{C}(\infty,u_{2})$, and the positive \Levy\ copula is such that $\mathcal{C}(u_{1},\infty)=u_{1}$ and $\mathcal{C}(\infty,u_{2})=u_{2}$. 

In the same way as a distributional copula connects marginal distribution functions to a joint distribution function, the \Levy\ copula connects marginal tail integrals to a tail integral. For a two-dimensional \Levy\ process with positive jumps, the tail integral $U$ is defined as
\begin{equation}
U(x_{1},x_{2})=\left \{ \begin{array}{ccl}    0                                       & \mbox{if} &  x_{1}=\infty \quad \mbox{and/or} \quad x_{2}=\infty \\
                                              \nu([x_{1},\infty )\times [x_{2},\infty )) & \mbox{if} & (x_{1},x_{2}) \in [0,\infty )^{2} \backslash \{0\} \\
                                              \infty                                  & \mbox{if} & (x_{1},x_{2})=0 
                         \end{array} \right.           
\end{equation}
and the marginal tail integrals are defined as
\begin{equation}
U_{1}(x_{1})=U(x_{1},0) \quad \mbox{and} \quad U_{2}(x_{2})=U(0,x_{2}).
\end{equation}
The following theorem due to \citet{cont04} is a version of Sklar's theorem for \Levy\ copulas.  

\newtheorem{levy_copula}{Theorem}
\begin{levy_copula}

Let $(S_{1},S_{2})$ be a two-dimensional \Levy\ process with positive jumps having tail integral $U$ and marginal tail integrals $U_{1}$ and $U_{2}$. There exists a two-dimensional positive \Levy\ copula $\mathcal{C}$ which characterizes the dependence structure of $(S_{1},S_{2})$, that is, for all $x_{1}$, $x_{2} \in [0,\infty]$,
\begin{equation}
U(x_{1},x_{2})=\mathcal{C}(U_{1}(x_{1}),U_{2}(x_{2})).
\label{levy_connect}
\end{equation}
If $U_{1}$ and $U_{2}$ are continuous, this \Levy\ copula is unique. Otherwise it is unique on $\mathrm{Ran} U_{1} \times \mathrm{Ran} U_{2}$. 

Conversely, let $S_{1}$ and $S_{2}$ be two one-dimensional \Levy\ processes with positive jumps having tail integrals $U_{1}$ and $U_{2}$ and let $\mathcal{C}$ be
a two-dimensional positive \Levy\ copula. Then there exsists a two-dimensional \Levy\ process with \Levy\ copula $\mathcal{C}$ and marginal tail integrals $U_{1}$ and $U_{2}$. Its tail integral is given by Eq.\ (\ref{levy_connect}).
\end{levy_copula}

The definition of the tail integral and its marginal tail integrals imply that $U_{1}(0)=\infty$ and $U_{2}(0)=\infty$. The singularity at zero is necessary to correctly account for jumps of the independent parts $S_{1}^{\perp}$ and $S_{2}^{\perp}$. Consider, for example, on the one hand
\begin{equation}
\mathcal{C}(U_{1}(x_{1}),U_{2}(0))=\mathcal{C}(U_{1}(x_{1}),\infty)=U_{1}(x_{1})=\nu([x_{1},\infty )\times [0,\infty)) \quad \mbox{for} \quad 0<x_{1}<\infty
\label{tail1}
\end{equation}
and, on the other hand
\begin{equation}
\lim_{x_{2} \downarrow 0} \mathcal{C}(U_{1}(x_{1}),U_{2}(x_{2}))=\lim_{x_{2} \downarrow 0} U(x_{1},x_{2})= \lim_{x_{2} \downarrow 0} 
\nu([x_{1},\infty)\times [x_{2},\infty)) \quad \mbox{for} \quad 
0 \le x_{1} < \infty.
\label{tail2}
\end{equation}
The difference 
\begin{equation}
\nu([x_{1},\infty )\times [0,\infty))-\lim_{x_{2} \downarrow 0} \nu([x_{1},\infty)\times [x_{2},\infty))=\nu([x_{1},\infty) \times \{0\}) 
\label{tail_diff}
\end{equation}
for $0 < x_{1} < \infty$ between the tail integrals of Eqs.\ (\ref{tail1}) and (\ref{tail2}) corresponds to the tail integral of $S_{1}^{\perp}$. If $U_{2}(0)$ would have been defined as $\lim_{x_{2} \downarrow 0}U_{2}(x_{2})$, the difference of the tail integrals vanishes and $S_{1}^{\perp}$ does not jump almost surely.      

\subsection{Construction of positive \Levy\ copulas}

In case of independent $S_{1}$ and $S_{2}$, the \Levy\ measure is concentrated on the set $\{ x \in [0,\infty)^{2}: x_{1}x_{2}=0 \}$ and the tail integral takes the form
\begin{equation}
U(x_{1},x_{2})=U_{1}(x_{1})1_{x_{2}=0}+U_{2}(x_{2})1_{x_{1}=0}.
\end{equation}
With the help of Eq.\ (\ref{levy_connect}), we find that the independence \Levy\ copula $\mathcal{C}_{\perp}$ is given by
\begin{equation}
\mathcal{C}_{\perp}(u_{1},u_{2})=u_{1}1_{u_{2}=\infty}+u_{2}1_{u_{1}=\infty}.
\end{equation} 
In case of comonotonic $S_{1}$ and $S_{2}$, the \Levy\ measure is concentrated on an increasing set $\mathcal{S}$ and the tail integral takes the form
\begin{equation}
U(x_{1},x_{2})=\min(U_{1}(x_{1}),U_{2}(x_{2})),
\end{equation}     
which implies that the comonotonic \Levy\ copula $\mathcal{C}_{\parallel}$ is given by
\begin{equation}
\mathcal{C}_{\parallel}(u_{1},u_{2})=\min(u_{1},u_{2}).
\end{equation}

A \Levy\ copula $\mathcal{C}$ with a dependence that is between $\mathcal{C}_{\perp}$ and $\mathcal{C}_{\parallel}$ can be constructed in several ways, such as by an approach similar to the construction of Archimedean distributional copulas \citep{cont04}. 
Given a strictly decreasing convex function $\phi: [0,\infty] \rightarrow [0,\infty]$ such that $\phi(0)=\infty$ and $\phi(\infty)=0$, a positive two-dimensional Archimedean \Levy\ copula is defined as
\begin{equation}
\mathcal{C}(u_{1},u_{2})=\phi^{-1}(\phi(u_{1})+\phi(u_{2})).
\end{equation}
For $\phi(u)=u^{-\delta}$ with $\delta>0$, one obtains the Clayton \Levy\ copula
\begin{equation}
\mathcal{C}(u_{1},u_{2})=(u_{1}^{-\delta}+u_{2}^{-\delta})^{-1/\delta}.
\label{clayton}
\end{equation}
The Clayton \Levy\ copula includes the independence \Levy\ copula $\mathcal{C}_{\perp}$ and the comonotonic \Levy\ copula $\mathcal{C}_{\parallel}$ in the limits $\delta \downarrow 0$ and $\delta \rightarrow \infty$, respectively. 

\subsection{Dependence implied by the \Levy\ copula}

The bivariate compound Poisson process is fully determined by the \Levy\ measures $\nu_{1}^{\perp}$, $\nu_{2}^{\perp}$ and $\nu^{\parallel}$. Given a \Levy\ copula $\mathcal{C}$, these measures can be expressed in terms of the \Levy\ measures $\nu_{1}$ and $\nu_{2}$. All \Levy\ measures are defined in the same way as in Section \ref{CPP}, but now $\nu$ is concentrated on $[0,\infty)^{2}\backslash \{0\}$. The relation between the measures $\nu_{1}^{\perp}$, $\nu_{2}^{\perp}$ and $\nu^{\parallel}$ on the one hand, and the measures $\nu_{1}$ and $\nu_{2}$ on the other, is established in Appendix \ref{app_levy_copula}. In this Section, for later purposes, the relation between the measures is expressed in terms of frequencies and jump size distributions. 

The frequencies $\lambda_{1}^{\perp}$ and $\lambda_{2}^{\perp}$ of, respectively, $S_{1}^{\perp}$ and $S_{2}^{\perp}$ are given by
\begin{equation}
\lambda_{1}^{\perp}=\nu_{1}^{\perp}((0,\infty))=\lambda_{1}-\lambda^{\parallel} \quad \mbox{and} \quad
\lambda_{2}^{\perp}=\nu_{2}^{\perp}((0,\infty))=\lambda_{2}-\lambda^{\parallel},
\end{equation}
where  
\begin{equation}
\lambda_{1}=\nu_{1}((0,\infty)), \quad \lambda_{2}=\nu_{2}((0,\infty)) \quad \mbox{and} \quad \lambda^{\parallel}=\nu^{\parallel}((0,\infty)\times(0,\infty))
\end{equation}
denote the frequencies of, respectively, $S_{1}$, $S_{2}$ and $S^{\parallel}$. In terms of the \Levy\ copula, $\lambda^{\parallel}$ takes the form
\begin{equation}
\lambda^{\parallel}=\mathcal{C}(\lambda_{1},\lambda_{2}).
\end{equation}
The survival function $\bar{F}_{1}^{\perp}$ of $\Delta S_{1}^{\perp}$ is defined as
\begin{equation}
\bar{F}_{1}^{\perp}(x_{1})=\mathbb{P}(\Delta S_{1}^{\perp} \in (x_{1},\infty))=\frac{\mathbb{P}(Y_{i} \in (x_{1},\infty) \times \{0\})}{\mathbb{P}(Y_{i} \in (0,\infty) \times \{0\})}=\frac{\nu_{1}^{\perp}((x_{1},\infty))}{\lambda_{1}^{\perp}},
\end{equation}
where $0 \le x_{1} < \infty$. In terms of the \Levy\ copula, $\bar{F}_{1}^{\perp}$ takes the form 
\begin{equation}
\bar{F}_{1}^{\perp}(x_{1})=\frac{\lambda_{1}}{\lambda_{1}^{\perp}}\bar{F}_{1}(x_{1})-\frac{1}{\lambda_{1}^{\perp}}\mathcal{C}(\lambda_{1}\bar{F}_{1}(x_{1}),\lambda_{2}),
\label{F1_perp}
\end{equation}
where $\bar{F}_{1}$ denotes the survival function of $\Delta S_{1}$. The distribution function $F_{1}^{\perp}$ of $\Delta S_{1}^{\perp}$ is related to $\bar{F}_{1}^{\perp}$ by
\begin{equation}
F_{1}^{\perp}(x_{1})=\mathbb{P}(\Delta S_{1}^{\perp} \in [0,x_{1}])=1-\bar{F}_{1}^{\perp}(x_{1}).
\label{F1_survival}
\end{equation}
Similarly, the distribution function $F_{1}$ of $\Delta S_{1}$ is given by $F_{1}=1-\bar{F}_{1}$. At this point, we have related $F_{1}^{\perp}$ to $\mathcal{C}$, $\lambda_{1}$, $\lambda_{2}$ and $F_{1}$. In the same way, the distribution function $F_{2}^{\perp}$ of $\Delta S_{2}^{\perp}$ is related to $\mathcal{C}$, $\lambda_{1}$, $\lambda_{2}$ and the distribution function $F_{2}$ of $\Delta S_{2}$. In terms of the survival functions
\begin{equation}
\bar{F}_{2}^{\perp}=1-F_{2}^{\perp} \quad \mbox{and} \quad \bar{F}_{2}=1-F_{2}, 
\label{F2_survival}
\end{equation}
the relation takes the form
\begin{equation}
\bar{F}_{2}^{\perp}(x_{2})=\frac{\lambda_{2}}{\lambda_{2}^{\perp}}\bar{F}_{2}(x_{2})-\frac{1}{\lambda_{2}^{\perp}}\mathcal{C}(\lambda_{1},\lambda_{2}\bar{F}_{2}(x_{2})).
\label{F2_perp}
\end{equation}
Finally, the joint survival function $\bar{F}^{\parallel}$ of $\Delta S^{\parallel}$ is defined as 
\begin{equation}
\begin{split}
\bar{F}^{\parallel}(x_{1},x_{2}) & =\mathbb{P}(\Delta S_{1}^{\parallel} \in (x_{1},\infty),\Delta S_{2}^{\parallel} \in (x_{2},\infty)) \\
 & = \frac{\mathbb{P}(Y_{i} \in (x_{1},\infty) \times (x_{2},\infty))}{\mathbb{P}(Y_{i} \in (0,\infty)^{2})}=
\frac{\nu^{\parallel}((x_{1},\infty)\times(x_{2},\infty))}{\lambda^{\parallel}},
\end{split}
\end{equation}
where $0 \le x_{1},x_{2} < \infty$. In terms of the \Levy\ copula, $\bar{F}^{\parallel}$ takes the form
\begin{equation}
\bar{F}^{\parallel}(x_{1},x_{2})=\frac{1}{\lambda^{\parallel}}\mathcal{C}(\lambda_{1}\bar{F}_{1}(x_{1}),\lambda_{2}\bar{F}_{2}(x_{2})).
\label{F_parallel}
\end{equation}
The distribution function $F^{\parallel}$ of $\Delta S^{\parallel}$ is related to $\bar{F}^{\parallel}$ by
\begin{equation}
F^{\parallel}(x_{1},x_{2}) = 1-\bar{F}^{\parallel}(x_{1},0)-\bar{F}^{\parallel}(0,x_{2})+\bar{F}^{\parallel}(x_{1},x_{2}).    
\label{2_survival}     
\end{equation}

\subsection{Examples of implied dependence}

A distributional survival copula $\bar{C}$ of $\bar{F}^{\parallel}$ satisfies
\begin{equation}
\bar{F}^{\parallel}(x_{1},x_{2})=\bar{C}(\bar{F}_{1}^{\parallel}(x_{1}),\bar{F}_{2}^{\parallel}(x_{2}))
\label{dist_copula}
\end{equation} 
where  
\begin{equation}
\bar{F}_{1}^{\parallel}(x_{1})=\bar{F}^{\parallel}(x_{1},0) \quad \mbox{and} \quad \bar{F}_{2}^{\parallel}(x_{2})=\bar{F}^{\parallel}(0,x_{2}).
\end{equation}
We assume here that $\bar{F}_{1}$ and $\bar{F}_{2}$ are continuous, which implies that $\bar{C}$ is unique. For the Clayton \Levy\ copula (\ref{clayton}), substituting 
\begin{multline}
\bar{F}_{1}^{\parallel}(x_{1})=\left(\frac{(\lambda_{1}\bar{F}_{1}(x_{1}))^{-\delta}+\lambda_{2}^{-\delta}}{\lambda_{1}^{-\delta}+\lambda_{2}^{-\delta}}\right)^{-1/\delta} \equiv u_{1} \quad \mbox{and} \\
\bar{F}_{2}^{\parallel}(x_{2})=\left(\frac{(\lambda_{2}\bar{F}_{2}(x_{2}))^{-\delta}+\lambda_{1}^{-\delta}}{\lambda_{1}^{-\delta}+\lambda_{2}^{-\delta}}\right)^{-1/\delta} \equiv u_{2}
\end{multline}
in the left- and right-hand side of Eq.\ (\ref{dist_copula}) gives
\begin{equation}
\bar{C}(u_{1},u_{2})=(u_{1}^{-\delta}+u_{2}^{-\delta}-1)^{-1/\delta},
\end{equation}
which is the distributional Clayton copula. The distributional copula $C$ of $F^{\parallel}$ takes the form
\begin{equation}
C(u_{1},u_{2})=((1-u_{1})^{-\delta}+(1-u_{2})^{-\delta}-1)^{-1/\delta}+u_{1}+u_{2}-1,
\label{dist_cop}
\end{equation}
which collapses to $uv$ for $\delta \downarrow 0$ and to $\min(u,v)$ for $\delta \rightarrow \infty$. The frequency $\lambda^{\parallel}$ implied by the Clayton \Levy\ copula takes the form
\begin{equation}
\lambda^{\parallel}=(\lambda_{1}^{-\delta}+\lambda_{2}^{-\delta})^{-1/\delta},
\end{equation}
which collapses to zero for $\delta \downarrow 0$ and to $\min(\lambda_{1},\lambda_{2})$ for $\delta \rightarrow \infty$. In summary, for $\delta \downarrow 0$, the Clayton \Levy\ copula implies $\lambda^{\parallel}=0$ and independent components of $S^{\parallel}$, while, for $\delta \rightarrow \infty$, it implies  $\lambda^{\parallel}=\min(\lambda_{1},\lambda_{2})$ and comonotonic components of $S^{\parallel}$.  

As a second example, we consider the pure common shock \Levy\ copula defined by \citet{ava12} as
\begin{equation}
\mathcal{C}(u_{1},u_{2})=\delta u_{1}u_{2} 1_{\max(u_{1},u_{2})<\infty} + u_{1} 1_{u_{2}=\infty} +u_{2} 1_{u_{1}=\infty}, \quad \mbox{where} \quad
0 \le \delta \le \min\left(\frac{1}{\lambda_{1}},\frac{1}{\lambda_{2}}\right).
\end{equation}
For the pure common shock \Levy\ copula, substituting
\begin{equation}
\bar{F}_{1}^{\parallel}(x_{1}) \equiv u_{1} \quad \mbox{and} \quad \bar{F}_{2}^{\parallel}(x_{2}) \equiv u_{2}
\end{equation}
in the left- and right-hand side of Eq.\ (\ref{dist_copula}) gives
\begin{equation}
\bar{C}(u_{1},u_{2})=u_{1}u_{2},
\end{equation}
which implies that the components of $S^{\parallel}$ are independent. The frequency $\lambda^{\parallel}$ implied by the pure common shock \Levy\ copula takes the form
\begin{equation}
\lambda^{\parallel}=\delta \lambda_{1} \lambda_{2},
\end{equation}
which equals zero if $\delta=0$ and $\min(\lambda_{1},\lambda_{2})$ if $\delta=\min(1/\lambda_{1},1/\lambda_{2})$. In summary, the dependence implied by the pure common shock \Levy\ copula is between frequencies only.

\section{Observation scheme and maximum likelihood estimation}
\label{MLE}

We consider a sample of jumps of a positive bivariate compound Poisson process discretely observed up to time $t=T$. The observation scheme is such that all jump sizes are observed, but it is not known which jumps stem from common shocks. We consider a partition 
of $[0,T]$ in $M$ intervals of equal length. The partition is chosen such that jumps of separate intervals can realistically be assumed not to stem from common shocks. In the context of operational risk modelling, with $T/M$ either being a month or a quarter, this is the observation scheme typically assumed in the advanced measurement approach. 

The objective of this work is to estimate the parameters of the \Levy\ copula in the observation scheme described above.
A possible solution is to construct a likelihood function based on all possible combinations of jumps within each interval. If, within a certain interval, there are $n_{1}$ jumps within loss category one and $n_{2}$ jumps within loss category two, one can distinguish between $\min(n_{1},n_{2})+1$ possibilities for the number of common jumps $0 \le n^{\parallel} \le \min(n_{1},n_{2})$. Given a certain $n^{\parallel}$, there are 
\begin{equation}
\frac{(\max(n_{1},n_{2}))!}{(\max(n_{1},n_{2})-n^{\parallel})!}\frac{(\min(n_{1},n_{2}))!}{(\min(n_{1},n_{2})-n^{\parallel})!}
\nonumber
\end{equation}        
possibilities of distributing the common jumps over the observed jump sizes. Due to the large number of possibilities, a likelihood function based on all combinations of jumps is not feasible. An alternative approach is to construct a likelihood function based on the number of jumps and the expected jump sizes within the intervals. This approach, however, is also not feasible because the convolutions involved typically have no closed-form expressions. In the method proposed here, we use
a sample consisting of the number of jumps and the maximum jump sizes within the intervals. For such a sample, we derive a closed-form likelihood function. Alternatively, a closed-form likelihood function based on the minimum jump sizes can also derived. In the context of operational risk modelling, however, one can expect the likelihood function based on maximum losses to me more variable with respect to model parameters than the likelihood function based on minimum losses.       

\subsection{Discrete processes for maximum likelihood estimation} 

We consider a partition $\Pi=\{t_{0},t_{1},..,t_{M}\}$ of $[0,T]$, where $t_{0}=0$, $t_{M}=T$ and $t_{i}-t_{i-1}=T/M$ for all $i=1,\ldots,M$. The discrete process 
$Z=(Z_{i})_{i=1,\ldots,M}=(Z_{i1},Z_{i2})_{i=1,\ldots,M}$ is defined as
\begin{equation}
Z_{ij}=\max_{t_{i-1}<s \le t_{i}}\Delta S_{j}(s)=\max(Z_{ij}^{\perp},Z_{ij}^{\parallel}),
\end{equation}
where 
\begin{equation}
Z_{ij}^{\perp}=\max_{t_{i-1} < s \le t_{i}} \Delta S_{j}^{\perp}(s) \quad \mbox{and} \quad Z_{ij}^{\parallel}=\max_{t_{i-1} < s \le t_{i}} \Delta S_{j}^{\parallel}(s).
\end{equation}
The random vector $Z_{i}$ is independent of $\mathcal{F}(t_{i-1})$ and has the same distribution for all $i$. In the same way as $Z$, we define the discrete processes $Z^{\perp}$ and $Z^{\parallel}$ based on $Z_{ij}^{\perp}$ and $Z_{ij}^{\parallel}$, respectively. These processes have the same properties as $Z$, but, in contrast to $Z$, they cannot be observed. The processes $Z^{\perp}$ and $Z^{\parallel}$ are independent. Also, the components of $Z^{\perp}$ are independent.
   
The discrete process $\tilde{N}=(\tilde{N}_{i})_{i=1,\ldots,M}=(\tilde{N}_{i1},\tilde{N}_{i2})_{i=1,\ldots,M}$ is defined as
\begin{equation}
\tilde{N}_{ij}=N_{j}(t_{i})-N_{j}(t_{i-1}),
\end{equation}
where the continuous process $N_{j}$ counts the number of jumps of $S_{j}$. The random variable $\tilde{N}_{ij}$ has decomposition
\begin{equation}
\tilde{N}_{ij}=\tilde{N}_{ij}^{\perp}+\tilde{N}_{i}^{\parallel},
\end{equation}
where
\begin{equation}
\tilde{N}_{ij}^{\perp}=N_{j}^{\perp}(t_{i})-N_{j}^{\perp}(t_{i-1}) \quad \mbox{and} \quad 
\tilde{N}_{i}^{\parallel}=N^{\parallel}(t_{i})-N^{\parallel}(t_{i-1}).
\end{equation}
Here, the continuous processes $N_{j}^{\perp}$ and $N^{\parallel}$ count the number of jumps of $S^{\perp}_{j}$ and $S^{\parallel}$, respectively.  
The random vector $\tilde{N}_{i}$ is independent of $\mathcal{F}(t_{i-1})$ and has the same distribution for all $i$. In the same way as $\tilde{N}$, we define the discrete process $\tilde{N}^{\perp}$ based on $\tilde{N}_{ij}^{\perp}$. The process $\tilde{N}^{\parallel}=(\tilde{N}_{i}^{\parallel})_{i=1,\ldots,M}$ is the discrete process corresponding to $N^{\parallel}$. The processes $\tilde{N}^{\perp}$ and $\tilde{N}^{\parallel}$ have the same properties as $\tilde{N}$, but, in contrast to $\tilde{N}$, they cannot be observed. The processes $\tilde{N}^{\perp}$ and $\tilde{N}^{\parallel}$ are independent. Also, the components of $\tilde{N}^{\perp}$ are independent.

\subsection{The likelihood function}

Realizations of the process $Z$ are collected in an $M \times 2$ matrix $z$ such that $z_{ij}$ is a realization of $Z_{ij}$. Similarly, the $M \times 2$ matrix $\tilde{n}$ holds realizations of $\tilde{N}$. The likelihood function $L_{z,\tilde{n}}$ corresponding to $z$ and $\tilde{n}$ takes the form
\begin{equation}
L_{z,\tilde{n}}=\prod_{i=1}^{M} \mathcal{L}_{z_{i1},z_{i2},\tilde{n}_{i1},\tilde{n}_{i2}},
\label{ll_prod}
\end{equation}
where $\mathcal{L}_{z_{i1},z_{i2},\tilde{n}_{i1},\tilde{n}_{i2}}$ denotes the likelihood function of the $i$-th row of $z$ and $\tilde{n}$. The entries of $L$ and $\mathcal{L}$ consist of $\lambda_{1}$, $\lambda_{2}$, the parameters of $F_{1}$, the parameters of $F_{2}$ and, finally, the parameters of the \Levy\ copula $\mathcal{C}$. The function $\mathcal{L}$ in Eq.\ (\ref{ll_prod}) has no discrete time label $i$ because the distributions of the random vectors do not depend on time. The likelihood function $L$ factorizes cf.\ (\ref{ll_prod}) because random vectors of different discrete time points are independent. The function $\mathcal{L}$ takes the form
\begin{equation}
\begin{split}
\mathcal{L}_{x,y,k,l} & = 1_{\{k=0,l=0\}} F_{0,0}(\infty,\infty) 
             + 1_{\{k>0,l=0\}} \frac{\partial F_{k,0}(x,\infty)}{\partial x} \\
            & + 1_{\{k=0,l>0\}} \frac{\partial F_{0,l}(\infty,y)}{\partial y} 
             + 1_{\{k>0,l>0\}} \frac{\partial^{2} F_{k,l}(x,y)}{\partial x \partial y}.                
\end{split}
\label{L_def}
\end{equation}
Here, the function $F_{k,l}$ is defined as
\begin{equation}
\begin{split}
F_{k,l}(x,y) = & \mathbb{P}(Z_{11}\le x,Z_{12}\le y,\tilde{N}_{11}=k,\tilde{N}_{12}=l) \\
                   = & \sum_{n^{\parallel}=0}^{\min(k,l)}
                    \mathbb{P}(Z_{11}\le x,Z_{12} \le y|\tilde{N}_{11}^{\perp}=k-n^{\parallel},\tilde{N}_{12}^{\perp}=l-n^{\parallel},\tilde{N}^{\parallel}_{1}
                    =n^{\parallel}) \times, \\
                  & \mathbb{P}(\tilde{N}_{11}^{\perp}=k-n^{\parallel},\tilde{N}_{12}^{\perp}=l-n^{\parallel},\tilde{N}^{\parallel}=n^{\parallel}),  
\end{split}                    
\label{F_def}
\end{equation}
where we have used that
\begin{equation}
\{\tilde{N}_{11}=k,\tilde{N}_{12}=l\}=\cup_{n^{\parallel}=0}^{\min(k,l)} \{\tilde{N}_{11}^{\perp}=k-n^{\parallel},\tilde{N}_{12}^{\perp}=l-n^{\parallel},\tilde{N}^{\parallel}=n^{\parallel}\}.
\end{equation}
The functions $F_{k,0}$ and $F_{0,l}$ with $k,l > 0$ are assumed to be differentiable with respect to the first and second entry, respectively. The functions $F_{k,l}$ with $k,l > 0$ are assumed to be continuously differentiable of second order. As we will see later, these assumptions hold if $F_{1}$ and $F_{2}$ are continuously differentiable and the \Levy\ copula $\mathcal{C}$ is continuously differentiable of second order on $(0,\lambda_{1})\times (0,\lambda_{2})$. 

To find an explicit expression for $\mathcal{L}$, we need to calculate $\mathbb{P}(\tilde{N}_{11}^{\perp}=k-n^{\parallel},\tilde{N}_{12}^{\perp}=l-n^{\parallel},\tilde{N}^{\parallel}_{1}=n^{\parallel})$ and to calculate the conditional probability appearing in Eq.\ (\ref{F_def}). Because $\tilde{N}_{11}^{\perp}$, $\tilde{N}_{12}^{\perp}$ and $\tilde{N}^{\parallel}_{1}$ are statistically independent Poisson random variables with frequencies $\lambda_{1}^{\perp}T/M$, $\lambda_{2}^{\perp}T/M$ and $\lambda^{\parallel}T/M$, respectively, we find 
\begin{multline}
\mathbb{P}(\tilde{N}_{11}^{\perp}=k-n^{\parallel},\tilde{N}_{12}^{\perp}=l-n^{\parallel},\tilde{N}^{\parallel}_{1}=n^{\parallel})= \\ \frac{(\lambda_{1}^{\perp}T/M)^{k-n^{\parallel}}e^{-\lambda_{1}^{\perp}T/M}}{(k-n^{\parallel})!} 
\frac{(\lambda_{2}^{\perp}T/M)^{l-n^{\parallel}}e^{-\lambda_{2}^{\perp}T/M}}{(l-n^{\parallel})!} 
\frac{(\lambda^{\parallel}T/M)^{n^{\parallel}}e^{-\lambda^{\parallel}T/M}}{n^{\parallel}!}. 
\end{multline}
The conditional probability appearing in $F_{0,0}$ is given by
\begin{equation}
\mathbb{P}(Z_{11} \le x,Z_{12} \le y|\tilde{N}_{11}=0,\tilde{N}_{12}=0,\tilde{N}^{\parallel}_{1}=0)=1.
\label{pcon_00}
\end{equation}
The conditional probability appearing in $F_{k,0}$ with $k>0$ takes the form 
\begin{equation}
\mathbb{P}(Z_{11} \le x, Z_{12} \le y | \tilde{N}_{11}^{\perp}=k,\tilde{N}_{12}^{\perp}=0, \tilde{N}^{\parallel}_{1}=0)=\mathbb{P}(Z_{11}^{\perp} \le x | \tilde{N}_{11}^{\perp}=k)
\label{case2}
\end{equation}
where we have used  
\begin{equation}
\{ Z_{11} \le x,Z_{12} \le y \} \cap \{\tilde{N}_{11}^{\perp}=k,\tilde{N}_{12}^{\perp}=0, \tilde{N}^{\parallel}_{1}=0 \} = \{ Z_{11}^{\perp} \le x,\tilde{N}_{11}^{\perp}=k,\tilde{N}_{12}^{\perp}=0,\tilde{N}^{\parallel}_{1}=0 \}
\end{equation} 
and the independence of $\tilde{N}_{12}^{\perp}$ and $\tilde{N}^{\parallel}_{1}$ from $Z_{11}^{\perp}$ and $\tilde{N}_{11}^{\perp}$. In terms of the distribution function $F_{1}^{\perp}$ of $\Delta S_{1}^{\perp}$ defined by Eqs.\ (\ref{F1_perp}) and (\ref{F1_survival}), Eq.\ (\ref{case2}) becomes 
\begin{equation}
\mathbb{P}(Z_{11} \le x, Z_{12} \le y | \tilde{N}_{11}^{\perp}=k,\tilde{N}_{12}^{\perp}=0, \tilde{N}^{\parallel}_{1}=0)=(F_{1}^{\perp}(x))^{k},
\label{pcon_k0}
\end{equation}
where we have used that the jump sizes of $S_{1}^{\perp}$ are iid. The conditional probability appearing in $F_{0,l}$ with $l>0$ is treated similarly. 
Finally, we consider a conditional probability appearing in $F_{k,l}$ with $k,l>0$. Making use of the independence of the processes $S_{1}^{\perp}$, $S_{2}^{\perp}$ and $S^{\parallel}$, we find
\begin{equation}
\begin{split}
 & \mathbb{P}(Z_{11}\le x,Z_{12}\le y|\tilde{N}_{11}^{\perp}=k-n^{\parallel},\tilde{N}_{12}^{\perp}=l-n^{\parallel},\tilde{N}^{\parallel}_{1}=n^{\parallel}) \\
 & =\mathbb{P}(Z_{11}^{\perp}\le x,Z_{11}^{\parallel}\le x,Z_{12}^{\perp} \le y,Z_{12}^{\parallel} \le y|\tilde{N}_{11}^{\perp}=k-n^{\parallel},\tilde{N}_{12}^{\perp}=l-n^{\parallel},\tilde{N}^{\parallel}_{1}=n^{\parallel}) \\
 & =\mathbb{P}(Z_{11}^{\perp} \le x|\tilde{N}_{11}^{\perp}=k-n^{\parallel})\mathbb{P}(Z_{12}^{\perp} \le y|\tilde{N}_{12}^{\perp}=l-n^{\parallel})
    \mathbb{P}(Z_{11}^{\parallel} \le x,Z_{12}^{\parallel} \le y|\tilde{N}^{\parallel}_{1}=n^{\parallel}) \\
    & = (F_{1}^{\perp}(x))^{k-n^{\parallel}}(F_{2}^{\perp}(y))^{l-n^{\parallel}}(F^{\parallel}(x,y))^{n^{\parallel}},  
\end{split}
\label{pcon_kl}
\end{equation}
where $F_{2}^{\perp}$ is the distribution function of $\Delta S_{2}^{\perp}$ defined by Eqs.\ (\ref{F2_survival}) and (\ref{F2_perp}), and $F^{\parallel}$ is the distribution function of $\Delta S^{\parallel}$ defined by Eqs.\ (\ref{F_parallel}) and (\ref{2_survival}). 

Substituting the conditional probabilities of Eqs.\ (\ref{pcon_00},\ref{pcon_k0},\ref{pcon_kl}) in Eq.\ (\ref{F_def}) and subsequently substituting the resulting expression of $F_{k,l}$ in Eq.\ (\ref{L_def}), gives 
\begin{equation}
\begin{split}
\mathcal{L}_{x,y,k,l} & = 1_{\{k=0,l=0\}} \mathbb{P}(\tilde{N}_{11}^{\perp}=0,\tilde{N}_{12}^{\perp}=0,\tilde{N}^{\parallel}_{1}=0) \\
            & + 1_{\{k>0,l=0\}} \mathbb{P}(\tilde{N}_{11}^{\perp}=k,\tilde{N}_{12}^{\perp}=0,\tilde{N}^{\parallel}_{1}=0) k(F_{1}^{\perp}(x))^{k-1}f_{1}^{\perp}(x) \\
            & + 1_{\{k=0,l>0\}} \mathbb{P}(\tilde{N}_{11}^{\perp}=0,\tilde{N}_{12}^{\perp}=l,\tilde{N}^{\parallel}_{1}=0) l(F_{2}^{\perp}(y))^{l-1}f_{2}^{\perp}(y) \\
            & + 1_{\{k>0,l>0\}} \sum_{n^{\parallel}=0}^{\min(k,l)} 
                \mathbb{P}(\tilde{N}_{11}^{\perp}=k-n^{\parallel},\tilde{N}_{12}^{\perp}=l-n^{\parallel},\tilde{N}^{\parallel}_{1}=n^{\parallel}) \times \\ 
            &   \qquad \frac{\partial^{2}}{\partial x \partial y} \left[                    
                (F_{1}^{\perp}(x))^{k-n^{\parallel}}(F_{2}^{\perp}(y))^{l-n^{\parallel}}(F^{\parallel}(x,y))^{n^{\parallel}}            
                \right],
\end{split}
\end{equation}
where $f_{1}^{\perp}$ and $f_{2}^{\perp}$
are the probability density functions corresponding to, respectively, $F_{1}^{\perp}$ and $F_{2}^{\perp}$. From Eqs.\ (\ref{F1_perp}) and (\ref{F1_survival}), we find 
\begin{equation}
f_{1}^{\perp}(x)=\frac{d F_{1}^{\perp}(x)}{dx}=\frac{\lambda_{1}f_{1}(x)}{\lambda_{1}^{\perp}}\left[ 1- \left. \frac{\partial \mathcal{C}(u,\lambda_{2})}{\partial u}    \right|_{u=\lambda_{1}\bar{F}_{1}(x)} \right],
\end{equation}
where $f_{1}$ is the probability density function corresponding to $F_{1}$ and $\bar{F}_{1}=1-F_{1}$ is the survival function of $\Delta S_{1}$. The density $f_{1}^{\perp}$ exists because $F_{1}$ and $\mathcal{C}$ are differentiable. Similarly, from Eqs.\ (\ref{F2_survival}) and (\ref{F2_perp}), we find
\begin{equation}
f_{2}^{\perp}(y)=\frac{d F_{2}^{\perp}(y)}{dy}=\frac{\lambda_{2}f_{2}(y)}{\lambda_{2}^{\perp}}\left[ 1- \left. \frac{\partial \mathcal{C}(\lambda_{1},v)}{\partial v}    \right|_{v=\lambda_{2}\bar{F}_{2}(y)} \right],
\end{equation}
where $f_{2}$ is the probability density function corresponding to $F_{2}$ and $\bar{F}_{2}=1-F_{2}$ is the survival function of $\Delta S_{2}$.
The double derivative in the likelihood takes the form
\begin{equation}
\begin{split}
& \frac{\partial^{2}}{\partial x \partial y} \left[ (F_{1}^{\perp}(x))^{k-n^{\parallel}}(F_{2}^{\perp}(y))^{l-n^{\parallel}}(F^{\parallel}(x,y))^{n^{\parallel}} \right] \\ 
& = (k-n^{\parallel})(l-n^{\parallel})f_{1}^{\perp}(x)f_{2}^{\perp}(y)(F_{1}^{\perp}(x))^{k-n^{\parallel}-1}
(F_{2}^{\perp}(y))^{l-n^{\parallel}-1}(F^{\parallel}(x,y))^{n^{\parallel}} \\
& + n^{\parallel}(l-n^{\parallel})f_{2}^{\perp}(y)F_{1}^{\parallel}(x,y) 
(F_{1}^{\perp}(x))^{k-n^{\parallel}}(F_{2}^{\perp}(y))^{l-n^{\parallel}-1}(F^{\parallel}(x,y))^{n^{\parallel}-1} \\
& + n^{\parallel}(k-n^{\parallel})f_{1}^{\perp}(x)F_{2}^{\parallel}(x,y) 
(F_{2}^{\perp}(y))^{l-n^{\parallel}}(F_{1}^{\perp}(x))^{k-n^{\parallel}-1}(F^{\parallel}(x,y))^{n^{\parallel}-1} \\
& + n^{\parallel}(n^{\parallel}-1)F_{1}^{\parallel}(x,y)F_{2}^{\parallel}(x,y)(F_{1}^{\perp}(x))^{k-n^{\parallel}}
(F_{2}^{\perp}(y))^{l-n^{\parallel}}(F^{\parallel}(x,y))^{n^{\parallel}-2} \\
& + n^{\parallel}f^{\parallel}(x,y)(F_{1}^{\perp}(x))^{k-n^{\parallel}}
(F_{2}^{\perp}(y))^{l-n^{\parallel}}(F^{\parallel}(x,y))^{n^{\parallel}-1},
\end{split}
\label{double}
\end{equation}
where
\begin{equation}
F_{1}^{\parallel}(x,y)=\frac{\partial F^{\parallel}(x,y)}{\partial x}, \quad
F_{2}^{\parallel}(x,y)=\frac{\partial F^{\parallel}(x,y)}{\partial y}, \quad \mbox{and} \quad
f^{\parallel}(x,y)=\frac{\partial^{2}F^{\parallel}(x,y)}{\partial x \partial y}.
\label{partials}
\end{equation}
From Eqs.\ (\ref{F_parallel}) and (\ref{2_survival}), we find 
\begin{equation}
F_{1}^{\parallel}(x,y)=\frac{\lambda_{1}f_{1}(x)}{\lambda^{\parallel}} 
\left[ \left. \left(-
\frac{\partial \mathcal{C}(u,\lambda_{2}\bar{F}_{2}(y))}{\partial u} + \frac{\partial \mathcal{C}(u,\lambda_{2})}{\partial u}
\right) \right|_{u=\lambda_{1}\bar{F}_{1}(x)} \right].
\end{equation}
Similarly, $F_{2}^{\parallel}(x,y)$ takes the form
\begin{equation}
F_{2}^{\parallel}(x,y)=\frac{\lambda_{2}f_{2}(y)}{\lambda^{\parallel}} 
\left[ \left. \left(-
\frac{\partial \mathcal{C}(\lambda_{1}\bar{F}_{1}(x),v)}{\partial v} + \frac{\partial \mathcal{C}(\lambda_{1},v)}{\partial v}
\right) \right|_{v=\lambda_{2}\bar{F}_{2}(y)} \right].
\end{equation}
Finally, the density $f^{\parallel}$ is given by
\begin{equation}
f^{\parallel}(x,y)=\frac{\partial^{2}F^{\parallel}(x,y)}{\partial x \partial y} = 
\frac{\lambda_{1}\lambda_{2}f_{1}(x)f_{2}(y)}{\lambda^{\parallel}} 
\left. 
\frac{\partial^{2}\mathcal{C}(u,v)}{\partial u \partial v}
\right|_{u=\lambda_{1}\bar{F}_{1}(x),v=\lambda_{2}\bar{F}_{2}(y)}.
\end{equation}
The double derivative of Eq.\ (\ref{double}) is continuous if $F_{1}$ and $F_{2}$ are continuously differentiable, and the \Levy\ copula $\mathcal{C}$ is continuously differentiable of second order on $(0,\lambda_{1})\times (0,\lambda_{2})$. The likelihood function $\mathcal{L}$ (and thereby $L$) is now completely specified in terms of the marginal frequencies, the marginal jump size distribution functions and the \Levy\ copula. In the limit of continuous observation ($M \rightarrow \infty$), the likelihood $L$ converges to the likelihood derived by \citet{Esm01}. 

\subsection{Maximizing the likelihood function}

The parameters of the process $S$ can be estimated by maximizing the likelihood function $L$ with respect to all its entries simultaneously. Alternatively, the likelihood function can be maximized with an approach similar to the inference functions for margins (IFM) approach of distributional copulas \citep{joe96}. The IFM method consists of two steps. In the first step, the parameters of 
$S_{j}$ with $j=1,2$ are estimated by maximizing the marginal likelihood function 
\begin{equation}
\mathcal{H}_{j,s_{j}}=(\lambda_{j})^{N_{j}(T)}e^{-\lambda_{j}T} \prod_{l=1}^{N_{j}(T)} f_{j}((s_{j})_{l}),
\end{equation} 
where $s_{j}$ is the vector of jumps of $S_{j}$ within $[0,T]$. In the second step, the estimates of the marginal parameters are substituted in $L$ and the resulting likelihood function is maximized with respect to the \Levy\ copula parameters. The IFM approach seems particularly suitable in the observation scheme of this work because the method makes use of all jump sizes (rather than the maximum jump sizes and the number of jumps in the $M$ intervals) in estimating the marginal parameters.

\section{A simulation study}
\label{sim_study}

The quality of the estimation method of Section \ref{MLE} is tested in a bootstrap analysis. The analysis consists of sampling many times from $S$ on a period $[0,1]$ and estimating its parameters. The marginal jump size distribution function $F_{j}$ with $j=1,2$ is given by  
\begin{equation}
F_{j}(x)=1-e^{-\theta_{j}x} 
\end{equation}
and dependence is introduced by the Clayton \Levy\ copula of Eq.\ (\ref{clayton}).
The process $S$ is thus described by five parameters (the marginal frequencies $\lambda_{1}$ and $\lambda_{2}$, the parameters $\theta_{1}$ and $\theta_{2}$ of the marginal jump size distribution functions and the parameter $\delta$ of the \Levy\ copula). Given the parameters of the process, the sampling algorithm consists of the following steps:
\begin{itemize}
\item{Draw $N_{1}^{\perp}$ and $N_{2}^{\perp}$ from a Poisson distribution with frequency $\lambda_{1}^{\perp}$ and $\lambda_{2}^{\perp}$, respectively.}
\item{Draw $N^{\parallel}$ from a Poisson distribution with frequency $\lambda^{\parallel}$.}
\item{Draw $N_{1}^{\perp}$ times from a uniform $[0,1]$ distribution. The resulting draws are the jump times of $S_{1}^{\perp}$. The $N_{2}^{\perp}$ jump times of 
$S_{2}^{\perp}$ are determined similarly.}  
\item{Draw $N_{1}^{\perp}$ times from a uniform $[0,1]$ distribution and apply the inverse of $F_{1}^{\perp}$ to each draw. The resulting numbers are the jump sizes of $S_{1}^{\perp}$. The jump sizes of $S_{2}^{\perp}$ are determined similarly.}
\item{Draw $N^{\parallel}$ times from a uniform $[0,1]$ distribution and apply the inverse of the marginal distribution function $F_{1}^{\parallel}$ defined as $F_{1}^{\parallel}(x)=F^{\parallel}(x,\infty)$ to each draw. The resulting numbers $x_{i}$ with $i=1,\ldots,N^{\parallel}$ are the jump sizes of $S_{1}^{\parallel}$. (Note that the marginal distribution function $F_{1}^{\parallel}$ defined here has one entry. In contrast, the function $F_{1}^{\parallel}$ defined in Eq.\ (\ref{partials}) with two entries denotes the partial derivative of $F^{\parallel}$ with respect to the first entry. We will use $F_{1}^{\parallel}$ to denote both functions. The number of entries indicates to which function it refers.)}
\item{Draw $N^{\parallel}$ times from a uniform $[0,1]$ distribution. The resulting draws are denoted by $u_{i}$ with $i=1,\dots,N^{\parallel}$. Apply the inverse of the distribution function $H_{x_{i}}(y)=F_{1}^{\parallel}(x_{i},y)/f_{1}^{\parallel}(x_{i})$, where $f_{1}^{\parallel}(x)=dF_{1}^{\parallel}(x)/dx$, to $u_{i}$ for all $i=1,\ldots,N^{\parallel}$. The resulting numbers $y_{i}$ with $i=1,\ldots,N^{\parallel}$ are the jump sizes of $S_{2}^{\parallel}$.} 
\end{itemize} 
In the last step, we have used that 
\begin{equation}
\begin{split}
\mathbb{P}(\Delta S^{\parallel}_{2} \le y| \Delta S_{1}^{\parallel} = x) & = \lim_{\Delta x \downarrow 0} \mathbb{P}(\Delta S^{\parallel}_{2} \le y| x \le \Delta S_{1}^{\parallel} \le x+\Delta x) \\
& = \lim_{\Delta x \downarrow 0} \frac{\mathbb{P}(x \le \Delta S_{1}^{\parallel} \le x+\Delta x, \Delta S^{\parallel}_{2} \le y)}{\mathbb{P}(x \le \Delta S_{1}^{\parallel} \le x+\Delta x)} \\
& = \lim_{\Delta x \downarrow 0} \frac{F^{\parallel}(x+\Delta x,y)-F^{\parallel}(x,y)}{F_{1}^{\parallel}(x+\Delta x)-F_{1}^{\parallel}(x)} = 
\frac{F_{1}^{\parallel}(x,y)}{f^{\parallel}_{1}(x)}=H_{x}(y). 
\end{split}
\end{equation} 
The algorithm to sample from $S$ is the same as the one used by \citet{Esm01} apart from the step where $N_{1}^{\perp}$ and $N_{2}^{\perp}$ are drawn. (
These authors first draw $N_{1}$, $N_{2}$ and $N^{\parallel}$ and then calculate $N_{1}^{\perp}$ and $N_{2}^{\perp}$ as $N_{1}^{\perp}=N_{1}-N^{\parallel}$ and $N_{2}^{\perp}=N_{2}-N^{\parallel}$. This method may accidentally work, but it could result in negative $N_{1}^{\perp}$ and $N_{2}^{\perp}$.)
To prepare a sample to which we can apply the methods of Section \ref{MLE}, we perform the following steps 
\begin{itemize}
\item{Divide the observation period $[0,1]$ in $M$ intervals of equal length and determine $Z_{i1}$ and $Z_{i2}$ for all $i=1,\ldots,M$. This results in an $M \times 2$ matrix $z$ of maximum jump sizes. Also determine $\tilde{N}_{i1}$ and $\tilde{N}_{i2}$ for all $i=1,\ldots,M$. This results in an $M \times 2$ matrix $\tilde{n}$ of number of jumps.}
\item{Determine the vector $s_{1}$ of all jump sizes of $S_{1}$ on $[0,1]$. Similarly, determine the vector $s_{2}$ of all jump sizes of $S_{2}$. (The vectors $s_{1}$ and $s_{2}$ are used in the IFM approach.)}    
\end{itemize} 

The algorithm described above is repeated many times and based on each $z$, $\tilde{n}$, $s_{1}$ and $s_{2}$, the parameters of $S$ are estimated. Statistics of the resulting bootstrap estimates are given in Tables \ref{results1} and \ref{results2} for the IFM approach and in Table \ref{results3} for maximum likelihood estimation. From Table \ref{results1}, we find that the difference between the bootstrap mean and the true value of $\delta$ is within a standard error of the mean (bootstrap standard deviation divided by 10). This indicates that the estimate of $\delta$ with the IFM approach is unbiased for the process under study and $M=100$. From Table \ref{results2}, we find that the estimate of $\delta$ remains unbiased for other values of $\delta$ and $M$. Also, for a fixed value of $M$, the bootstrap standard deviation is seen to be approximately proportional to $\delta$, which
means that the relative precision of the estimate of $\delta$ does not depend on $\delta$. 
From Table \ref{results3}, we find that the difference between the bootstrap mean and the true value of $\delta$ is within a standard error of the mean (bootstrap standard deviation divided by $\sqrt{50}$). This indicates that the maximum likelihood estimate of $\delta$ is unbiased for the process under study and $M=100$. Similarly, from Tables \ref{results1} and \ref{results3}, estimates of the parameters of the marginal processes are seen to be unbiased in both methods. In terms of the bootstrap standard deviations, however, the IFM approach provides estimates of a slightly better quality.

\begin{table}[ht!]
\caption{Parameter value, bootstrap mean and bootstrap standard deviation of the parameters of a bivariate compound Poisson process. The parameters consist of the marginal frequencies $\lambda_{1}$ and $\lambda_{2}$, the parameters $\theta_{1}$ and $\theta_{2}$ of the exponential marginal jump size distribution functions, and the parameter $\delta$ of the Clayton \Levy\ copula. The simulation period $[0,1]$ is divided in 100 intervals of equal length and the parameters are estimated with the IFM approach in 100 bootstrap samples.}  
\begin{center}
\begin{tabular}{lrrrrr}  \hline 
                                       & $\lambda_{1}$ & $\lambda_{2}$ & $\theta_{1}$ & $\theta_{2}$ & $\delta$   \\  \hline 
value                                  & 1000          & 1000          & 1            & 1            & 1          \\
bootstrap mean                         & 1004          & 999           & 0.999        & 1.002        & 1.007      \\
bootstrap standard deviation           & 33            & 36            & 0.032        & 0.031        & 0.114      \\  \hline
\end{tabular}
\end{center}
\label{results1}
\end{table} 

\begin{table}[ht!]
\caption{Parameter value, bootstrap mean and bootstrap standard deviation of the Clayton \Levy\ copula parameter $\delta$ of a bivariate compound Poisson process. The process has marginal frequencies $\lambda_{1}=\lambda_{2}=1000$ and marginal exponential jump size distribution functions with parameters $\theta_{1}=\theta_{2}=1$. The simulation period $[0,1]$ is divided in $M$ intervals of equal length and the parameter $\delta$ is estimated with the IFM approach in 100 bootstrap samples. For $\delta=1$, the same bootstrap samples are used as in Table \ref{results1}.}   
\begin{center}
\begin{tabular}{cccc} \hline
$M$   &  \multicolumn{3}{c}{$\delta$} \\ \cline{2-4}
      &  value      & bootstrap mean          & bootstrap standard deviation       \\
50    &  1          & 0.989                   & 0.143                              \\
50    &  5          & 4.999                   & 0.543                              \\
100   &  1          & 1.007                   & 0.114                              \\
100   &  5          & 5.006                   & 0.417                        \\ \hline                           
\end{tabular}    
\end{center}
\label{results2}
\end{table}

\begin{table}[ht!]
\caption{Parameter value, bootstrap mean and bootstrap standard deviation of the parameters of a bivariate compound Poisson process. The parameters consist of the marginal frequencies $\lambda_{1}$ and $\lambda_{2}$, the parameters $\theta_{1}$ and $\theta_{2}$ of the exponential marginal jump size distribution functions, and the parameter $\delta$ of the Clayton \Levy\ copula. The simulation period $[0,1]$ is divided in 100 intervals of equal length and the parameters are estimated with maximum likelihood estimation in 50 bootstrap samples.}  
\begin{center}
\begin{tabular}{lccccc}  \hline 
                                       & $\lambda_{1}$ & $\lambda_{2}$ & $\theta_{1}$ & $\theta_{2}$ & $\delta$   \\  \hline 
value                                  & 1000          & 1000          & 1            & 1            & 1          \\
bootstrap mean                         & 1004          & 997           & 0.993        & 0.994        & 0.992      \\
bootstrap standard deviation           & 37            & 39            & 0.036        & 0.042        & 0.113      \\  \hline
\end{tabular}
\end{center}
\label{results3}
\end{table}

\section{A real data analysis}
\label{real_data}

We apply the methodology developed in Section \ref{MLE} to the Danish fire loss data set publicly available at \url{http://www.ma.hw.ac.uk/~mcneil}. The data set consists of fire insurance data divided into loss of building, loss of content and loss of profit. Common shocks are known in this data set and a \Levy\ copula has already been fitted to the data based on the likelihood function of continuous observation \citep{Esm01}. To mimick an actual loss data set faced by banks in operational risk modelling, we consider a monthly observation and remove common shock information within months.

\subsection{Description of the data and pre-processing}

Details about the loss data set are described at its source \url{http://www.ma.hw.ac.uk/~mcneil}. The data set consists of 2167 fire loss events over the period 1980 up to 1990 (11 years). They have been adjusted for inflation and are given with respect to the year 1985 in millions of Danish Kroner. Each loss event is divided into a loss of building, a loss of content and a loss of profit. In order to make a comparison with the case of known common shocks studied by \citet{Esm01}, we apply the same pre-processing to the data. This means that the loss of profit is not taken into account because it rarely has a non-zero value. For the remaining two loss categories, we consider only losses exceeding a million Kroner and take the logarithm of these losses. The resulting data set consists of 940 transformed loss events. 

To prepare a sample for the discretely observed process under study in this work, we consider a monthly partition of the 11 years and determine the maximum loss and the number of losses for each month and each loss category. This results in a $132 \times 2$ matrix $z$ of maximum losses and a $132 \times 2$ matrix $\tilde{n}$ of number of losses. Details about the maximum losses and the number of losses are given in Figure 1. 
To prepare for the IFM approach, we construct the vectors $s_{1}$ and $s_{2}$ holding, respectively, all losses of building and all losses of content. The vector $s_{1}$ holds 782 losses and the vector $s_{2}$ holds 456 losses.   

\begin{figure}
\begin{center}
\includegraphics[height=20cm,width=15cm]{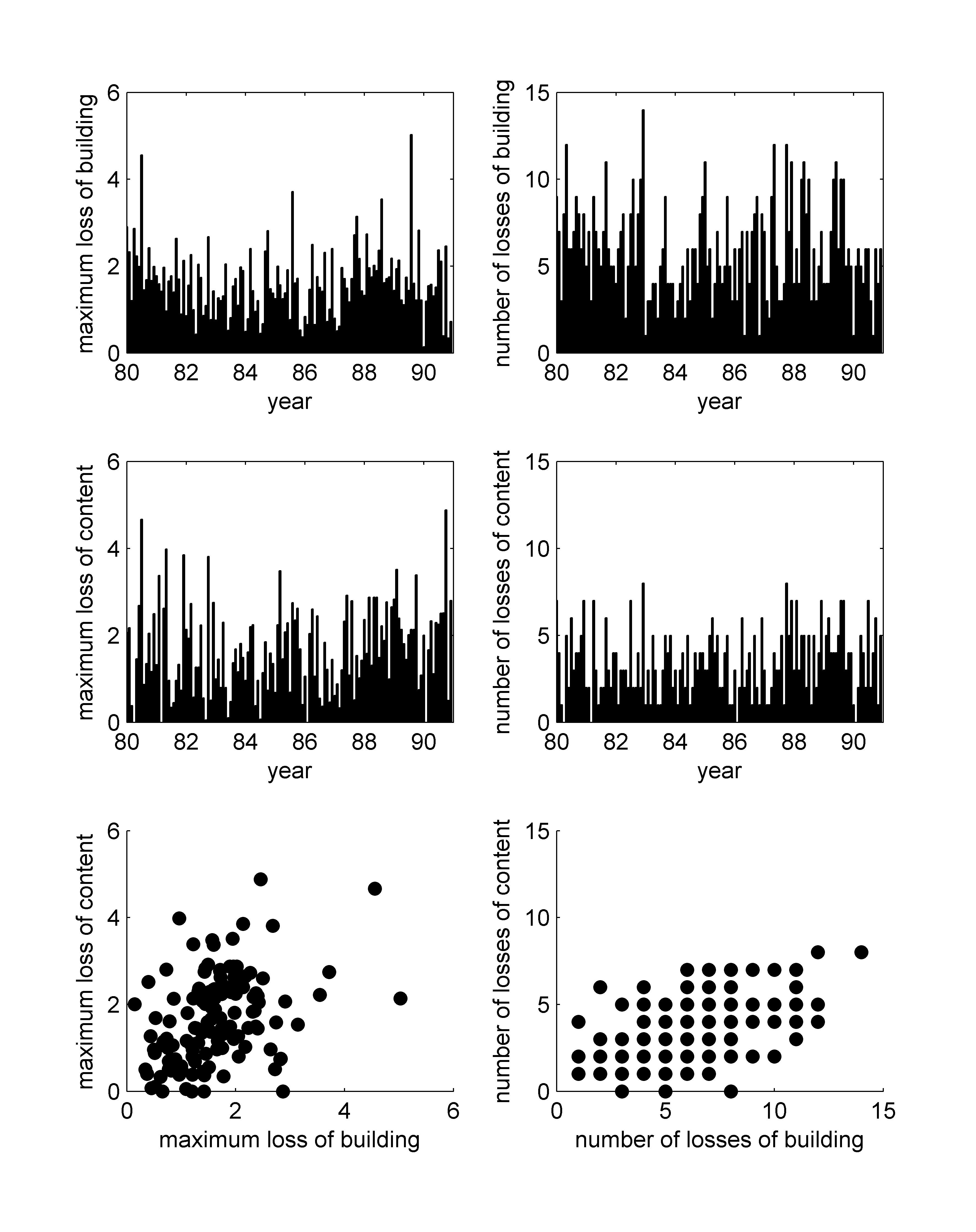}
\vspace{-1cm}
\caption{Monthly maximum log-losses and monthly number of losses for the Danish fire loss data set. The maximum log-losses are given as a function of time in the upper left panel (loss of building) and the middle left panel (loss of content). Similarly, the number of losses are given as a function of time in the upper right panel (loss of building) and the middle right panel (loss of content). The maximum log-losses of the two loss categories are plotted against each other in the lower left panel. Similarly, the number of losses of the two loss categories are plotted against each other in the lower right panel.}  
\end{center}
\label{sample_fig}
\end{figure}

\subsection{Marginal processes}

We will estimate a \Levy\ copula with the IFM approach. This means that the parameters of the process $S_{1}$ of loss of building and the parameters of the process $S_{2}$ of loss of content are based on, respectively, $s_{1}$ and $s_{2}$.
We use the same marginal jump size distributions as \citet{Esm01}. These authors use a Weibull distribution function for the log-losses and study the quality of the fit in terms of QQ-plots. The Weibull distribution function $F_{j}$ of $\Delta S_{j}$ takes the form
\begin{equation}
F_{j}(x)=1-e^{-(x/\alpha_{j})^{\beta_{j}}} 
\end{equation}
where $x \ge 0$. By maximizing the marginal likelihood function $\mathcal{H}_{j,s_{j}}$, we find estimates of $\lambda_{j}$, $\alpha_{j}$ and $\beta_{j}$ for $j=1,2$. The estimates are listed in Table \ref{results4}. 

\begin{table}[ht!]
\caption{Parameter estimates of the compound Poisson process corresponding to loss of building and the compound Poisson process corresponding to loss of content of the Danish fire loss data set. The annual frequency of losses of building is denoted by $\lambda_{1}$ and the annual frequency of losses of content by $\lambda_{2}$. The parameters $\alpha_{1}$ and $\beta_{1}$ correspond to the Weibull distribution function of the log-losses of building. Similarly, the parameters $\alpha_{2}$ and $\beta_{2}$ correspond to the Weibull distribution function of the log-losses of content.} 
\begin{center}
\begin{tabular}{ccccccc}  \hline 
             & $\lambda_{1}$  & $\lambda_{2}$ & $\alpha_{1}$    & $\beta_{1}$    & $\alpha_{2}$    & $\beta_{2}$     \\ \hline
estimate     & 71.1          & 41.5          & 0.818           & 1.197          & 1.036           & 1.131           \\  \hline
\end{tabular}
\end{center}
\label{results4}
\end{table}

\subsection{A goodness of fit test and selection of a \Levy\ copula}
\label{GoF}

In the case of continuous observation or known common shocks, one can check if a \Levy\ copula provides a reasonable fit to the data by inspecting the goodness of fit of the implied distributional copula between the components of $\Delta S^{\parallel}$. For a Clayton \Levy\ copula, for example, the implied distributional copula is given by Eq.\ (\ref{dist_cop}). The goodness of fit of a distributional copula can be assessed by transforming the correponding pseudo sample of probabilities to another sample of probabilities based on the copula \citep{brey10}. If the copula provides a reasonable fit, the transformed probabilities are realizations of statistically independent uniform $[0,1]$ random variables. This can be tested by applying the inverse of the standard normal distribution function to the $[0,1]$ random variables and subsequently performing standard statistical tests. 

In the discrete observation scheme discussed in this work, common shocks are unknown. In order to test the fit and select a reasonable \Levy\ copula, we construct a method similar to the goodness of fit test of distributional copulas. For this purpose, we define the distribution function $G_{k,l}$ as
\begin{equation}
G_{k,l}(x,y)=\mathbb{P}(Z_{11} \le x,Z_{12} \le y|\tilde{N}_{11}=k,\tilde{N}_{12}=l)=\frac{F_{k,l}(x,y)}{F_{k,l}(\infty,\infty)}.
\end{equation}
For the $i$-th time interval, the probability $u_{i1}$ is calculated as  
\begin{equation}
u_{i1}=G_{\tilde{n}_{i1},\tilde{n}_{i2}}(z_{i1},\infty).
\end{equation}
We calculate the distribution function $H_{x,k,l}(y)$ as
\begin{equation}
\begin{split}
H_{x,k,l}(y) & =\mathbb{P}(Z_{12} \le y|Z_{11}=x,\tilde{N}_{11}=k,\tilde{N}_{12}=l) \\
             & =\frac{\mathbb{P}(Z_{11}=x,Z_{12} \le y,\tilde{N}_{11}=k,\tilde{N}_{12}=l)}{\mathbb{P}(Z_{11}=x,\tilde{N}_{11}=k,\tilde{N}_{2}=l)} \\
             & = \lim_{\Delta x \downarrow 0} \frac{\mathbb{P}(x \le Z_{11} \le x+\Delta x,Z_{12} \le     
                  y,\tilde{N}_{11}=k,\tilde{N}_{12}=l)}{\mathbb{P}(x \le Z_{11} \le x+\Delta x,\tilde{N}_{11}=k,\tilde{N}_{12}=l)} \\                           
             & =\frac{\frac{\partial F_{k,l}(x,y)}{\partial x}}{\frac{\partial F_{k,l}(x,\infty)}{\partial x}}.
\end{split}
\end{equation}
The probability $u_{i2}$ is calculated as 
\begin{equation}
u_{i2}=H_{z_{i1},\tilde{n}_{i1},\tilde{n}_{i2}}(z_{i2}).
\end{equation}
We select the rows of $u$ that correspond to the rows of $\tilde{n}$ of which both elements are non-zero. These rows correspond to the time intervals in which at least one loss is recorded in both loss categories. The selected rows are collected in a matrix $v$ with two columns. In case of the Danish fire loss data, $v$ is a $128 \times 2$ matrix. If the data generating process is correctly specified, $v$ should be a realization of a random matrix with independent and uniform $[0,1]$ elements. To test this assumption, we translate $v$ into a matrix $w$ by
\begin{equation}
w_{ij}=\Phi^{-1}(v_{ij}), 
\end{equation}
where $\Phi$ denotes the standard normal distribution function. The matrix $w$ should have independent standard normal elements. For the Danish fire loss data, this is tested for the pure common shock \Levy\ copula and the Clayton \Levy\ copula. The results are listed in Table \ref{results5}. For the pure common shock \Levy\ copula, the estimated correlation coefficient $\hat{\rho}_{12}$ between the columns of $w$ deviates from zero at the 0.01 level. This indicates that the pure common shock \Levy\ copula is probably not appropriate for the data set. In contrast, the Clayton \Levy\ copula provides a good fit. This is in line with the results of \citet{Esm01}.         

\begin{table}[ht!]
\caption{Test results of the goodness of fit of a bivariate compound Poisson process parameterized with a \Levy\ copula to the monthly Danish fire loss data set. The test is based on a $128 \times 2$ matrix $w$ which, if the process is correctly specified, should contain realizations of statistically independent standard normal variables. The test results consist of the Jarque-Bera statistic ($\mbox{JB}$) of \citet{jar80}, the mean $\hat{\mu}$, the standard deviation $\hat{\sigma}$ and the serial correlation coefficient $\hat{\rho}$ of the two columns of $w$. A subscript indicates to which column the statistic refers. The test results also include the correlation coefficient $\hat{\rho}_{12}$ between the columns of $w$. The p-values corresponding to the null hypothesis that the elements of $w$ are realizations of independent standard normal variables, are given between parentheses below the corresponding statistics. The p-values are two-sided for $\hat{\mu}_{1}$, $\hat{\mu}_{2}$, $\hat{\rho}_{1}$, $\hat{\rho}_{2}$ and $\hat{\rho}_{12}$, and one-sided for $\hat{\sigma}_{1}$ and $\hat{\sigma}_{2}$.} 
\begin{center}
\resizebox{16cm}{!}{
\begin{tabular}{lccccccccc}  \hline 
\Levy\ copula        &  $\mbox{JB}_{1}$  &  $\mbox{JB}_{2}$ &  $\hat{\mu}_{1}$   &  $\hat{\mu}_{2}$  &  $\hat{\sigma}_{1}$  &  $\hat{\sigma}_{2}$  & $\hat{\rho}_{1}$  
& $\hat{\rho}_{2}$ & $\hat{\rho}_{12}$ \\  \hline     
pure common shock  &  0.79  &  0.60  & -0.01  & -0.01  &  1.04  &  0.95  &  0.09  &  0.05  &  0.25   \\
                   & (0.64) & (0.72) & (0.94) & (0.89) & (0.26) & (0.25) & (0.29) & (0.60) & (0.01)  \\
Clayton            &  1.14  &  0.41  & -0.02  & -0.01  &  1.00  &  1.05  &  0.08  &  0.09  & -0.09   \\
                   & (0.52) & (0.80) & (0.77) & (0.92) & (0.46) & (0.21) & (0.38) & (0.31) & (0.29)  \\ \hline                               
\end{tabular}}
\end{center}
\label{results5}
\end{table}   

\subsection{Estimation results}

Based on the analysis of Section \ref{GoF}, we use a Clayton \Levy\ copula to fit the monthly Danish fire loss data. The parameters of the process are estimated with the IFM approach and the results are listed in Table \ref{results6}. We also estimate $\delta$ with the IFM approach in case of known common shocks. This results in an estimate of 0.903, a bootstrap mean of 0.904 and a bootstrap standard deviation of 0.043. These bootstrap statistics are based on 100 bootstrap samples. As expected, the bootstrap standard deviation with unknown common shocks is larger than with known common shocks. The IFM estimate of 0.903 is close to the maximum likelihood estimate of 0.953 reported by \citet{Esm01}.             

\begin{table}[ht!]
\caption{Estimate, bootstrap mean and bootstrap standard deviation of the parameters of a bivariate compound Poisson process fitted to the monthly Danish fire loss data. The parameters consist of the marginal frequencies $\lambda_{1}$ (loss of building) and $\lambda_{2}$ (loss of content), the parameters $\alpha_{1}$ and $\beta_{1}$ of the Weibull distribution of log-losses of building, the parameters $\alpha_{2}$ and $\beta_{2}$ of the Weibull distribution of log-losses of content, and the parameter $\delta$ of the Clayton \Levy\ copula. The parameters are estimated with the IFM approach in 100 bootstrap samples.}  
\begin{center}
\begin{tabular}{lccccccc}  \hline 
                          & $\lambda_{1}$  & $\lambda_{2}$ & $\alpha_{1}$    & $\beta_{1}$    & $\alpha_{2}$    & $\beta_{2}$ & $\delta$    \\ \hline
estimate            & 71.1           & 41.5          & 0.818           & 1.197          & 1.036           & 1.131       & 0.695       \\ 
bootstrap mean            & 70.5           & 41.3          & 0.818           & 1.206          & 1.038           & 1.141       & 0.699       \\
bootstrap standard deviation    &  2.6           &  2.1          & 0.024           & 0.034          & 0.047           & 0.046       & 0.092       \\ \hline
\end{tabular}
\end{center}
\label{results6}
\end{table}

\section{Conclusions}
\label{conclusions}

In summary, we have developed a method to estimate a \Levy\ copula of a bivariate compound Poisson process in case the process is observed discretely with knowledge about all jump sizes, but without knowledge of which jumps stem from common shocks. The method is tested in a simulation study with a Clayton \Levy\ copula. The results indicate that the method is unbiased in small samples and that the bootstrap standard deviation of the Clayton \Levy\ copula parameter is approximately proportional to its bootstrap mean. A goodness of fit test for the \Levy\ copula is developed and applied to monthly log-losses of the Danish fire loss data set. The results indicate that the Clayton \Levy\ copula provides a good fit to the data set. 

The method developed in this work is particularly useful in the context of operational risk modelling in which common shocks are typically unknown. To model dependencies between operational losses of different loss categories, the common practice in the banking industry is to use a distributional copula between either the number of losses or the aggregate losses within a certain time window. A disadvantage of this approach is that the distributional copula depends non-trivially on the length of the time window. If one has, for example, estimated a distributional copula between monthly losses, the distributional copula between yearly losses is typically unknown. A second disadvantage of the approach is that the nature of the model depends on the level of granularity. If one combines, for example, two loss categories connected by a distributional copula, the new loss category is typically not compound Poisson. These two issues are resolved by a multivariate compound Poisson process, which can be parsimoniously modelled with a \Levy\ copula in a bottom-up approach.

\appendix

\section{Connection between \Levy\ measures}
\label{app_levy_copula}

In this Appendix, we relate the \Levy\ measures $\nu_{1}^{\perp}$, $\nu_{2}^{\perp}$ and $\nu^{\parallel}$ to the \Levy\ measures $\nu_{1}$, $\nu_{2}$ and the \Levy\ copula $\mathcal{C}$. 
On a Borel set $(x_{1},\infty)$ with $0 \le x_{1} < \infty$, the \Levy\ measure $\nu_{1}^{\perp}$ is given by
\begin{equation}
\nu^{\perp}_{1}((x_{1},\infty))=\nu((x_{1},\infty)\times \{0\}),
\end{equation}
which is equivalent to
\begin{equation}
\nu^{\perp}_{1}((x_{1},\infty))=\nu((x_{1},\infty) \times [0,\infty))-\lim_{x_{2} \downarrow 0}\nu((x_{1},\infty) \times [x_{2},\infty)).
\end{equation}
In terms of
\begin{equation}
\nu_{1}((x_{1},\infty)=\nu((x_{1},\infty)\times  [0,\infty)) 
\end{equation}
and the \Levy\ copula, $\nu_{1}^{\perp}((x_{1},\infty)$ takes the form 
\begin{equation}
\begin{split}
\nu_{1}^{\perp}((x_{1},\infty)) & =
\nu_{1}((x_{1},\infty))-\lim_{x'_{1} \downarrow x_{1}}\lim_{x_{2} \downarrow 0} \nu([x'_{1},\infty),[x_{2},\infty)) \\
& = \nu_{1}((x_{1},\infty))-\lim_{x'_{1} \downarrow x_{1}}\lim_{x_{2} \downarrow 0} \mathcal{C}(\nu_{1}([x'_{1},\infty),\nu_{2}([x_{2},\infty)) \\
                                & =\nu_{1}((x_{1},\infty))-\mathcal{C}(\nu_{1}((x_{1},\infty),\nu_{2}(0,\infty)),
\end{split}
\label{nu1perp}
\end{equation}
Similarly, on a Borel set $(x_{2},\infty)$ for $0\le x_{2}<\infty$, the measure $\nu_{2}^{\perp}$ takes the form
\begin{equation}
\nu_{2}^{\perp}((x_{2},\infty))=\nu_{2}((x_{2},\infty))-\mathcal{C}(\nu_{1}(0,\infty),\nu_{2}((x_{2},\infty))),
\label{nu2perp}
\end{equation}
Finally, on a Borel set $(x_{1},\infty)\times(x_{2},\infty)$ with $0 \le x_{1},x_{2} < \infty$, the measure $\nu^{\parallel}$ is given by
\begin{equation} 
\begin{split}
\nu^{\parallel}((x_{1},\infty)\times (x_{2},\infty)) & =\lim_{x'_{1} \downarrow x_{1}}\lim_{x'_{2} \downarrow x_{2}}
\nu^{\parallel}([x'_{1},\infty)\times [x'_{2},\infty)) \\
& = \lim_{x'_{1} \downarrow x_{1}}\lim_{x'_{2} \downarrow x_{2}}  \mathcal{C}(\nu_{1}([x'_{1},\infty),\nu_{2}[x'_{2},\infty)) \\
 & =\mathcal{C}(\nu_{1}((x_{1},\infty),\nu_{2}(x_{2},\infty)).
\end{split}
\label{nuparallel}
\end{equation}

\end{document}